\documentclass[aps,twocolumn,showpacs]{revtex4}
\usepackage{amsmath}
\usepackage{amsfonts}
\usepackage{graphicx}

\begin{document}

\title{Periodic giant-persistent current in sharp pulses on a ring}

\author{Y. Z. He}
\author{C. G. Bao}
\thanks{Corresponding author: stsbcg@mail.sysu.edu.cn}
\affiliation{
 Center of Theoretical Nuclear Physics,
 National Laboratory of Heavy Ion Collisions,
 Lanzhou, 730000.
 \\ and \\
 Department of Physics,
 Sun Yat-Sen University,
 Guangzhou, 510275, P.R. China
}

\begin{abstract}
We show here a mesoscopic device based on a narrow ring containing
an electron. In the device, an amount of energy is stored in
advance. Similar to the pendulum, an exact periodic motion of the
electron is thereby initiated afterward. The motion appears as a
series of sharp pulses, and is in nature different from the well
known Aharonov-Bohm (A-B) oscillation. In particular, the pulses of
current can be tuned to be very strong (say, more than two orders
stronger than the usual A-B current). Related theory and numerical
results are presented.
\end{abstract}
\pacs{73.23.Ra, 74.78.Na, 74.90.+n}

\maketitle

\section{Introduction}

It is well known that mesoscopic devices are not only interesting in
academic aspect but also promising in technical aspect. Among these
devices, the quantum ring is distinguished due to its special
geometry. Accordingly, special physical phenomena emerge, namely,
the normal and fractional Aharonov-Bohm (A-B) oscillations of the
ground state energy and persistent
current\cite{r_HU,r_KUF,r_MD,r_FA,r_RE,r_HAE,r_VS,r_HYZ}. The A-B
oscillation is caused by the variation of an external magnetic flux
threading the ring. At any instant, the current is uniform
(\textit{i.e.}, it is the same everywhere on the ring). This kind of
oscillation has already been extensively studied. In this paper, we
report a new kind of persistent current which is a series of pulses
and is different from the A-B current in nature. They move on the
ring strictly periodically. Both the intensity and the period can be
tuned. In particular, giant current (\textit{i.e.}, two or more
orders stronger than the A-B current) can be obtained.

It is recalled that the oldest clock is based on the periodic motion
of a classical pendulum, where the motion is initiated by an amount
of potential energy stored in advance. Similarly, in a mesoscopic
device, if an amount of energy has been stored in advance and can be
transformed to kinetic energy afterward, and if dissipation can be
neglected, a persistent periodic motion might exist based on the
principle of the pendulum. In this paper, an idea is proposed so
that an amount of energy can be stored in a system with an electron
moving on a super-conducting ring, and an exact periodic motion of
the electron can be thereby induced. Related theory and numerical
results are as follows.

\section{Hamiltonian and related theory}

We consider a very narrow ring which is a quasi-one-dimensional
system containing a free electron. A constant magnetic field is
preset vertical to the plane of the ring. A square-well potential
lying on a segment of the ring (say, caused by electrodes) is
applied to localize the electron. The Hamiltonian reads
\begin{equation}
 H^{\Phi }
 =G(-i\frac{\partial }{\partial \theta }+\Phi )^{2}+U(\theta )
 \label{e_H_Phi}
\end{equation}
where $\theta $ is the azimuthal angle of the electron, $G=\hbar
^{2}/(2m^{\ast }R^{2})$, $m^{\ast }$ is the effective mass, and $R$
is the radius of the ring. $\Phi $ is the magnetic flux threading
the ring in the unit $\Phi _{0}=hc/e$. $U(\theta )=0$ if $\pi -d\leq
\theta \leq \pi +d$, or $=U_{o}$ otherwise. $U_{o}$ is a positive
number and $2d$ is the width of the square-well. The potential is
deep enough so that at least a localized state is contained. The
electron is assumed to be in the ground state $\Psi _{g}$ with
energy $E_{g}>0$ initially. Thus it is localized in the beginning.

Suddenly, the well is removed ($U(\theta )$ becomes zero everywhere)
and at the same time the flux $\Phi $ is changed to $\Phi ^{\prime
}$. Accordingly, the Hamiltonian is suddenly changed from $H^{\Phi
}$ to
\begin{equation}
 H_{evol}
 =G(-i\frac{\partial }{\partial \theta }+\Phi ^{\prime })^{2}
 \label{e_H_evol}
\end{equation}
Obviously, $\Psi _{g}$ is no more the eigen-state of the new
Hamiltonian. Thus it begins to evolve.

Starting from $\Psi _{g}$, the formal time-dependent solution of the
Schr\"{o}dinger equation reads
\begin{equation}
 \Psi (t)=\exp(-iH_{evol}t/\hbar )\Psi _{g}
 \label{e_Psi_H_evol}
\end{equation}
which can be written in an applicable form when we know all the
eigen-states of $H_{evol}$. This is easy for one-dimensional rings.
They read simply $|k\rangle =\exp(ik\theta)/\sqrt{2\pi }$, where $k$
include all integers from $ -\infty $ to $+\infty $. With them,
eq.~(\ref{e_Psi_H_evol}) can be rewritten as
\begin{equation}
 \Psi (t)
 =\sum_{k}\exp(-iE_{k}t/\hbar) |k\rangle \langle k|\Psi _{g}\rangle
 \label{e_Psi_E_k}
\end{equation}
where $E_{k}=G(k+\Phi ^{\prime })^{2}$.

In order to obtain $\Psi _{g}$, $H^{\Phi }$ is diagonalized by using
the set $|k\rangle $ as basis functions. For numerical calculation
the summation of $k$ in eq.~(\ref{e_Psi_E_k}) can be confined within
a range, say, from $ -I_{\Phi }-30$ to $-I_{\Phi }+30$, where
$I_{\Phi }$ is the integer closest to $\Phi $. This could provide
the accuracy with at least four effective figures in numerical
results.

The matrix element reads
\begin{eqnarray}
 &&\langle k'|H^{\Phi}|k\rangle= \nonumber \\
 &&\left\{
  \begin{array}{lc}
   G(k+\Phi)^2+U_o(1-d/\pi), & \mbox{if}\ k=k' \\
   -\frac{U_o}{\pi(k-k')}\cos[(k-k')\pi]\sin[(k-k')d],  & \mbox{else}
  \end{array}
  \right.
 \label{e_H_Psi_kp_k}
\end{eqnarray}
After the diagonalization the ground state energy $E_{g}$ and the
ground state $\Psi _{g}=\sum_{k}C_{k}^{g}|k\rangle $ can be
obtained. In particular, $\langle k|\Psi _{g}\rangle =C_{k}^{g}$ can
be known. There are two noticeable features:

(i) From eq.~(\ref{e_H_Psi_kp_k}), we have $\langle k^{\prime
}|H^{\Phi }|k\rangle =\langle k^{^{\prime }}-I|H^{\Phi
+I}|k-I\rangle $ where $I$ is an arbitrary integer. Therefore the
Hamiltonian $H^{\Phi +I}$ and $H^{\Phi }$ have exactly the same
ground state energy $E_{g}$. And the ground state of $ H^{\Phi +I}$
reads $\Psi _{g}^{\Phi +I}=\sum_{k}C_{k+I}^{g}|k\rangle $. It
implies that, once the cases with $0\leq \Phi \leq 1$ are clear, the
cases with $\Phi <0$ and $\Phi >1$ are also clear.

(ii) When $\Phi $ is an integer or a half-integer, we have $\langle
k^{\prime }|H^{\Phi }|k\rangle =\langle -k^{\prime }-2\Phi |H^{\Phi
}|-k-2\Phi \rangle $. This equality would impose a kind of symmetry
between the coefficients $C_{k}^{g}$ and $C_{-k-2\Phi }^{g}$. Since
the ground state should not have a node at $\theta =\pi $, it can be
deduced that $ C_{k}^{g}=C_{-k-2\Phi }^{g}$ (if $\Phi $ is an
integer) or $ C_{k}^{g}=-C_{-k-2\Phi }^{g}$ (if $\Phi $ is a half
integer).

Let $t_{o}=\hbar /G$ be the unit of time, and $\tau =t/t_{o}$. From
eq.~(\ref {e_Psi_E_k}), we obtain the time-dependent density
\begin{eqnarray}
 \rho (\theta ,\tau )
 &=& |\Psi (t)|^{2} \nonumber \\
 &=& \frac{1}{2\pi }\sum_{k,k^{\prime }}C_{k}^{g}C_{k^{\prime }}^{g} \nonumber  \\
 & & \cos [(k-k^{\prime }) (\theta -(k+k^{\prime }+2\Phi ^{\prime})\tau )]
 \label{e_rho}
\end{eqnarray}
and, from the conservation of mass, the current
\begin{eqnarray}
 J(\theta ,\tau )
 &=& 2J_{A-B}\sum_{k,k^{\prime}}C_{k}^{g}C_{k^{\prime}}^{g}(k+\Phi ^{\prime }) \nonumber \\
 & & \cos [(k-k^{\prime }) (\theta -(k+k^{\prime }+2\Phi^{\prime})\tau )]
 \label{e_J}
\end{eqnarray}
where $J_{A-B}\equiv \hbar /4\pi m^{\ast }R^{2}$ is the maximal
current of the A-B oscillation of the one-electron ground state.
$J_{A-B}$ is served as a unit of current in the follows.

It was found from eqs.~(\ref{e_rho}) and~(\ref{e_J}) that the motion
of the electron would be periodic if $\Phi ^{\prime }$ satisfies the
following rule:

\textit{Let $I$, $I_{o}$ and $I_{e}$ denote positive integers, where
$I_{o}$ is an odd integer and $I_{e}$ even integer. (a) If
$(I_{e})_{\min }$ is the smallest even integer so that $|\Phi
^{\prime }|=I/(I_{e})_{\min }$, then the evolution is strictly
periodic with a period $\tau _{p}=(I_{e})_{\min }\pi $. (b) However,
for the special cases $|\Phi ^{\prime }|=I_{o}^{\prime }/2I_{o}$,
the period $\tau _{p}=I_{o}\pi $. (c) Otherwise, the evolution is
not strictly periodic.}

This rule implies that the periodicity is decisively determined by
$\Phi ^{\prime }$. For examples, when $\Phi ^{\prime }=1/8$, from
(a) $\tau _{p}=8\pi $; when $\Phi ^{\prime }=1/6=1/(2\times 3)$,
from (b) $\tau _{p}=3\pi $; when $\Phi ^{\prime }=1/5=2/(2\times
5)$, from (a) $\tau _{p}=10\pi $; and so on. In particular, when
$\Phi ^{\prime }=I=$ $2I/2$, we have from (a) $\tau _{p}=2\pi $
(this includes the case $\Phi ^{\prime }=0$ ). When $\Phi ^{\prime
}$ is a half-integer $I_{o}^{\prime }/2$, we have from (b) $\tau
_{p}=\pi $. Among all the choices of $\Phi ^{\prime }$, a
half-integer would lead to a shortest period.

Experimentally, in order to have an exact periodic motion, it is
clear from the rule that $\Phi ^{\prime }$ must be accurately given
without error. Otherwise, the motion is only approximately periodic.
The simplest way to meet this goal is just to set $\Phi ^{\prime
}=0$ (we shall study this case in detail). In order to obtain
numerical results, $m^{\ast }=0.063m_{e}$ (for a InGaAs ring),
$R=100nm$, $U_{o}=2G$, and $d=\pi /4$ are assumed. With these
choices, the time unit $t_{o}=1.09\times 10^{-11}\sec $, and the
period $2\pi t_{o}$ would be $6.85\times 10^{-11}\sec $. Evidently,
a larger $m^{\ast }$ and/or $R$ would lead to a longer period.
Numerical results of $ \rho $ and $J$ with $\Phi \geq 0$ and $\Phi
^{\prime }=0$ are given in the follows.

\section{Numerical results}

\begin{figure}[tbp]
 \centering \resizebox{0.95\columnwidth}{!}{
 \includegraphics{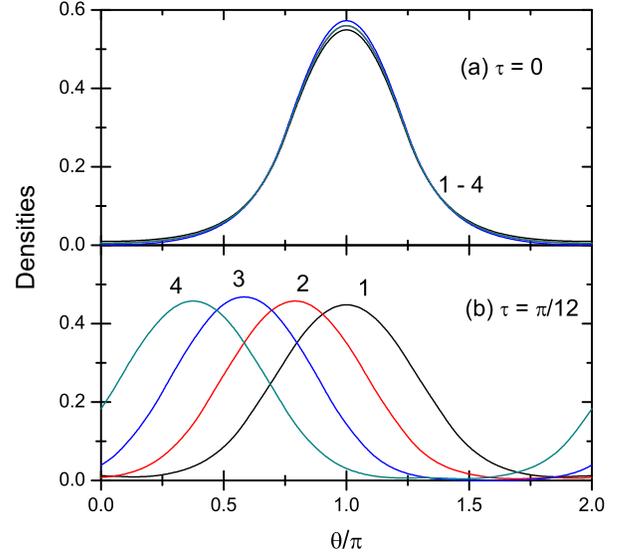} }
 \caption{(Color online). $\protect\rho (\protect\theta
,\protect\tau )$ plotted against $ \protect\theta $ when
$\protect\tau $ is given. $m^{\ast }=0.063m_{e}$ is assumed and the
ring has $R=100nm$. The preset potential has $U_{o}=2G$ and $
d=\protect\pi /4$. $\Phi^{\prime }$ is given zero. The above
parameters are also the same in following figures. The curves from
"1" to "4" have $\Phi =0$ , 1.25, 2.5, and 3.75, respectively.
$\protect\tau =0$ (a) and $\protect\pi /12$ (b). }
 \label{fig.1}
\end{figure}
\begin{figure}[tbp]
 \centering \resizebox{0.95\columnwidth}{!}{
 \includegraphics{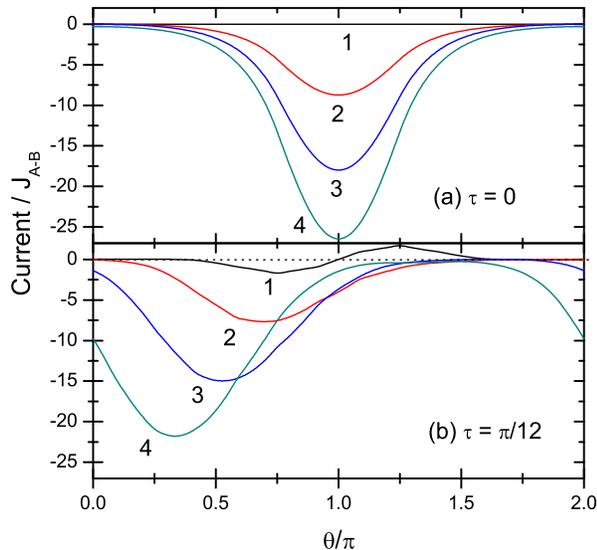} }
 \caption{(Color online). $J(\protect\theta
,\protect\tau )/J_{A-B}$ plotted against $\protect \theta $ when
$\protect\tau =0$ (a) and $\protect\tau =\protect\pi /12$ (b). The
curves from "1" to "4" have $\Phi =0$, 1.25, 2.5, and 3.75,
respectively, as in Fig.~\ref{fig.1}. $J_{A-B}\equiv \hbar
/(4\protect\pi m^{\ast }R^{2})$ is the maximal current of the A-B
oscillation. }
 \label{fig.2}
\end{figure}

How the density is distributed on the ring in the early stage of
evolution is shown in Fig.~\ref{fig.1}. When $\tau =0$, the electron
is localized around $\theta =\pi $ as shown in~\ref{fig.1}a as
mentioned, where the distribution is not sensitive to $\Phi $. When
the evolution starts, the peak of $\rho $ will shift left
(clockwise) as shown in~\ref{fig.1}b with $\tau =\pi /12$. However,
it would shift right if $\Phi <0$. The shift would be larger if
$\Phi $ is larger. The shift is caused by an initial current
$J(\theta ,0)$ shown in Fig.~\ref{fig.2}a. This current is created
simultaneously with the sudden creation of the new Hamiltonian. When
$\Phi =0$, $J(\theta ,0)=0$ (curve "1" of~\ref{fig.2}a). When $\Phi
\neq 0$, a larger $|\Phi |$ leads to a stronger $J(\theta ,0)$ as
shown by "2" to "4" of~\ref{fig.2}a, they are the motivity of the
evolution. In~\ref{fig.2}b, the current of "1" is either negative
(if $\theta <\pi $) or positive (if $\theta
>\pi $ ), it implies that the current flows to both sides if $\Phi
=0$. However, when $\Phi >0$, "2" to "4"\ of~\ref{fig.2}b are purely
negative implying going left. Obviously, a more negative current
leads to a larger shift of $\rho $ as previously shown
in~\ref{fig.1}b.

\begin{figure}[tbp]
 \centering \resizebox{0.95\columnwidth}{!}{
 \includegraphics{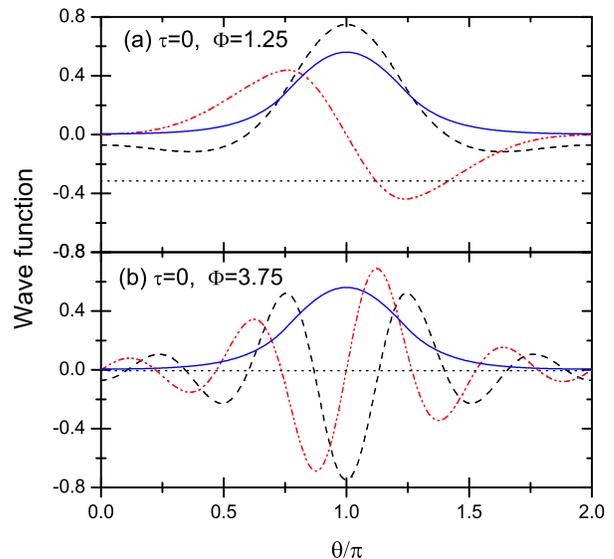} }
 \caption{(Color online). Wave function of the initial
state $\Psi _{g}(\protect\theta )$. Dash-line is for the real part
and dash-dot-dot-line is for the imaginary part, the solid line is
for $|\Psi _{g}|^{2}=\protect\rho (\protect\theta ,0) $. }
 \label{fig.3}
\end{figure}

To better understand the origin of $J(\theta ,0)$, we study the
structure of the initial state $\Psi _{g}$. It was found that $\Psi
_{g}$ is less sensitive to the parameters of the potential, but very
sensitive to $\Phi $. This fact is shown in Fig.~\ref{fig.3}.
Although $|\Psi _{g}|^{2}\equiv \rho (\theta ,0) $ plotted
in~\ref{fig.3}a and~\ref{fig.3}b (in solid lines) are nearly the
same disregarding the difference in $\Phi $, their wave functions
are greatly different. In~\ref{fig.3}b there are many nodes in both
the real and imaginary parts of $\Psi _{g}$. It is well known that
the number of nodes would be a measure of kinetic energy if the
magnetic field does not exist. Therefore, an amount of energy $
E_{init}$ has been stored in $\Psi _{g}$ initially via the preset
magnetic field. Once the field is canceled, the energy can be
released and motivates the evolution.

\begin{figure}[tbp]
 \centering \resizebox{0.95\columnwidth}{!}{
 \includegraphics{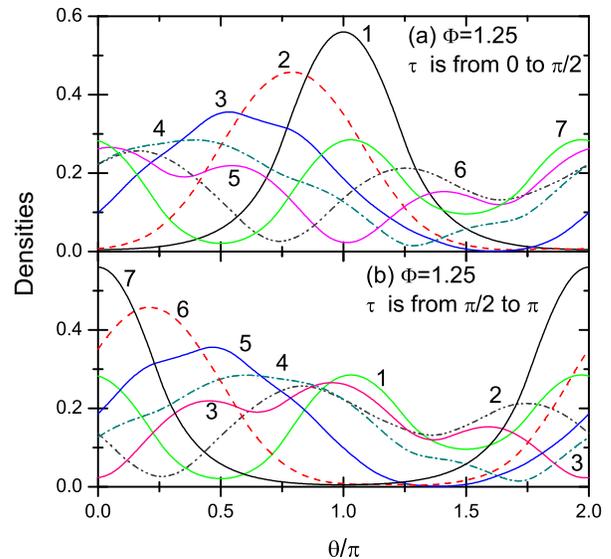} }
 \caption{(Color online). $\protect\rho (\protect\theta
,\protect\tau )$ plotted against $ \protect\theta $ when $\Phi
=1.25$ and $\protect\tau $ is given at a number of values. In a (b),
the curves "1" to "7" have $\protect\tau $ increases step by step
from 0 to $\protect\pi /2$ ($\protect\pi /2$ to $\protect\pi $ ),
each step the increase is $\protect\pi /12$. }
 \label{fig.4}
\end{figure}

The evolution of $\rho $\ is shown in Fig.~\ref{fig.4}, where $\Phi
=1.25$ and $\tau $ is from 0 to $\pi $. In the duration $[0,\pi
/4]$, the peak of $\rho $ starting from the initial end ($\theta
\approx \pi $) keeps going left (clockwise) and becomes broader and
broader as shown by the curves "1" to "4" of~\ref{fig.4}a. In this
duration, the distance that the peak has shifted increases
remarkably with $\Phi $. In~\ref{fig.4}a, the peak of "4" is apart
from the initial end by $\sim 5\pi /8$. However, if $\Phi =6.25$ for
an example, it would become $13\pi /8$. In the next duration $[\pi
/4,3\pi /4]$, $\rho $ becomes diffused and is widely distributed on
the ring with peaks and dips as shown by "5" to "7" of~\ref{fig.4}a
and "1" to "3" of~\ref{fig.4}b. Meanwhile, the classical picture of
motion is not clear. It can be called a duration of cruise. However,
when $\tau =\pi /2$, $\rho $ is nearly half-to-half concentrated in
both ends as shown by "7" of~\ref{fig.4}a (which is identical to "1"
of~\ref{fig.4}b) . When $ \tau =3\pi /4$, the diffused density
begins to be concentrated again and forms a broad main peak as shown
by "4" of~\ref{fig.4}b. The broad peak is roughly apart from the
opposite end ($\theta \approx 0$) by $5\pi /8$ (the same distance
mentioned above). Afterward, in $[3\pi /4,\pi ]$, the broad peak
keeps going left, becomes narrower, and arrives at the opposite end
exactly at $\tau =\pi $. This is shown by "4" to "7"
of~\ref{fig.4}b. Thus the evolution of $ \rho $ in the upper half
period can be divided into three stages, namely, shift- cruise
-shift. This qualitative feature is not affected by $\Phi $. In
particular, $\rho $ is peaked at the initial end when $\tau =0$,
peaked at the opposite end when $\tau =\pi $, \ and nearly
half-to-half concentrated in both ends when $\tau =\pi /2$
disregarding how $\Phi $ is. Nonetheless, how far the peak of $\rho
$ would shift within the durations $ [0,\pi /4]$ or $[3\pi /4,\pi ]$
depends seriously on $\Phi $. Besides, during the cruise, a larger
$\Phi $ leads to a stronger current in general.

When $\Phi ^{\prime }=0$ as we have chosen, it is straight forward
from eqs.~(\ref{e_rho}) and~(\ref{e_J}) to obtain $\rho (-\theta
,\pi +\tau )=\rho (\theta ,\pi -\tau )$ and $J(-\theta ,\pi +\tau
)=J(\theta ,\pi -\tau )$. Therefore, the evolution in $[\pi ,2\pi ]$
is just a time-inverse of the evolution in $[0,\pi ]$ together with
a spatial reflection against $\theta =0 $. It implies that, when
$\tau >\pi $, the peak would shift left from $ \theta =0$ to $-5\pi
/8$ (if $\Phi =1.25$), then a cruise, then a shift again from $\pi
+5\pi /8$ toward $\pi $. When $\tau =2\pi $, the peak arrives
exactly at the initial end $\theta =\pi $ and the period is ended,
and a new period will begin, and so on.

\begin{figure}[tbp]
 \centering \resizebox{0.95\columnwidth}{!}{
 \includegraphics{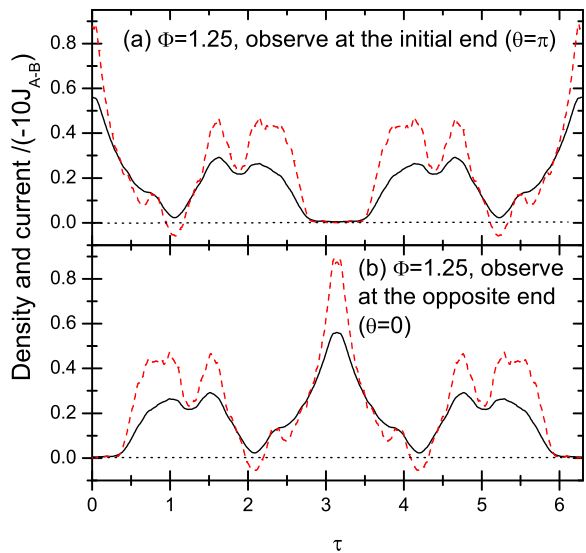} }
 \caption{(Color online). Density and current plotted
against $\protect\tau $ when $\Phi =1.25 $ and (a) $\protect\theta
=\protect\pi $ or (b) 0. The solid line is $ \protect\rho $, the
dash-line is $-0.1J/J_{A-B}$. The factor -0.1 is introduced to fit
the ordinate. }
 \label{fig.5}
\end{figure}

From Fig.~\ref{fig.4} we know that the density will highly
concentrated in the initial (opposite) end once and once when $\tau
=I_{e}\pi $ ($I_{o}\pi $). Accordingly, when one observe the
time-dependence of current at the initial (opposite) end, a series
of pulses will appear one-after-one as shown in Fig.~\ref{fig.5}.
For an example, there are five peaks in~\ref{fig.5}b at $\tau
=0.28\pi $, $ 0.49\pi $, $\pi $, $1.51\pi $, and $1.72\pi $,
respectively. At each of these instants, a pulse of current will
arrive at the opposite end. The pulse appearing at $\tau =\pi $ is
much stronger (associated with the curve "7" of~\ref{fig.4}b). There
are also five peaks in~\ref{fig.5}a, the highest peak is at $\tau
=0$ (associated with the curve "1" of~\ref{fig.4}a). Once the
strongest pulse has arrived at an end, the current at the other end
is zero simultaneously. The maximal current of the highest peak (the
strongest pulse) is found to be nearly linearly proportional to
$|\Phi |$. With the above parameters, we found an approximate
formula $J(0,\pi )\approx -6.9\Phi J_{A-B}$ (Say, when $\Phi =1.25$,
$J(0,\pi )=-8.6J_{A-B}$ as plotted in~\ref{fig.5}b. When $\Phi =30$,
it is more than 200 times stronger than the usual A-B current).
Thus, giant, punctual, and persistent sharp pulses of current can be
obtained. This is the most distinguished feature of the device.

The assumption that $\Phi ^{\prime }=0$ leads to a strict period
$\tau _{p}=2\pi $. When $\Phi ^{\prime }$ is nonzero, the periodic
behavior depends on $\Phi ^{\prime }$ extremely sensitively. For an
example, based on the rule, when $\Phi ^{\prime }=\frac{499}{1000}$,
we have $\tau _{p}=1000\pi $. However, when $\Phi ^{\prime
}=\frac{500}{1000}$, we have $ \tau _{p}=\pi $. That is, a very
slight change in $\Phi ^{\prime }$ might lead to a great change in
$\tau _{p}$. This high sensitivity is also a noticeable point.

\section{Conclusions}

In summary, a device with periodic persistent current in sharp
pulses is proposed based on a narrow ring. A crucial point is how to
store an amount of energy in advance. For this aim a potential and a
magnetic field is preset. Then, the potential is removed suddenly
and the flux jumps from $ \Phi $ to $\Phi ^{\prime }$
simultaneously. In this way, the amount of energy can be released
and motivate an exact periodic motion of the electron. The period is
controllable (say, by altering the materials, the size of the ring,
and/or $\Phi ^{\prime }$). On the contrary with the A-B current, the
present current is far from uniform but in sharp pulses, and can be
tuned to be very strong by increasing the difference $|\Phi ^{\prime
}-\Phi |$. The appearance of sharp pulses is a distinguished feature
of the device.

When $\Phi ^{\prime }=0$, the evolution in a period has been studied
in detail. Three stages (shift-cruise-shift) have been found. The
distance that the peak of $\rho $ has shifted during the stage of
"shift" depends seriously on $\Phi $, and a larger $|\Phi |$ will
lead to a longer distance. At the two ends of the ring sharp pulses
of current are found passing by one-after-one. The strongest pulses
will emerge at the initial (opposite) end each time when $\tau
=I_{e}\pi $ ($I_{o}\pi $). When $\Phi >15,$ the strongest current
$J(0,I_{o}\pi )$ is two or more orders stronger than the A-B
current.

When $\Phi ^{\prime }\neq 0$, the above periodic behavior will be
changed. When $\Phi ^{\prime }=I_{o}/2$, the period would be the
shortest $\pi t_{o}$ . It is noted that the current given by
eq.~(\ref{e_J}) contains the product $C_{k}^{g}(k+\Phi ^{\prime })$.
Due to the feature of the ground state, $ |C_{k}^{g}|$ would be
larger only if $k$ is close to $-\Phi $. Therefore, the contribution
of this product would be larger if $|\Phi ^{\prime }-\Phi |$ is
larger. This explains why a big jump in the flux is necessary to
produce the giant current.

Finally, it is noted that time-counting is important to developing
micro-techniques. The central element for time-counting would be a
device containing exact periodic motion. It is believed that,
comparing with smooth periodic motion, a device with sharp pulses
(refer to Fig.~\ref{fig.5}) might be a better candidate for this
purpose. Furthermore, due to the sharp pulses, the period $2\pi
t_{o}$\ might be accurately measured. This might lead to a better
understanding to the parameters of the system ($m^{\ast }$ and $R$).

\section*{Acknowledgments} The support from NSFC under the grant 10574163 and
10874249 is appreciated.

\end{document}